\documentclass[prb,floatfix,twocolumn,showpacs]{revtex4}
\usepackage{graphics}

\begin{document}

\title{Cherenkov radiation emitted by ultrafast laser pulses and the generation of coherent polaritons}
\author{J.~K.~Wahlstrand}
\author{R.~Merlin}
\affiliation{FOCUS Center and Department of Physics, The University of Michigan, Ann Arbor, MI 48109-1120}
\date{\today}

\begin{abstract}
We report on the generation of coherent phonon polaritons in ZnTe, GaP and
LiTaO$_{3}$ using ultrafast optical pulses. These polaritons are coupled
modes consisting of mostly far-infrared radiation and a small phonon
component, which are excited through nonlinear optical processes involving
the Raman and the second-order susceptibilities (difference frequency
generation). We probe their associated hybrid vibrational-electric field, in the THz range, by
electro-optic sampling methods. The measured field patterns agree very well
with calculations for the field due to a distribution of dipoles that
follows the shape and moves with the group velocity of the optical pulses.
For a tightly focused pulse, the pattern is
identical to that of classical Cherenkov radiation by a moving dipole.
Results for other shapes and, in particular, for the planar and
transient-grating geometries, are
accounted for by a convolution of the Cherenkov field due to a point dipole
with the function describing the slowly-varying intensity of the pulse.
Hence, polariton fields resulting from pulses of arbitrary shape can be
described quantitatively in terms of expressions for the Cherenkov radiation
emitted by an extended source. Using the Cherenkov approach, we recover the
phase-matching conditions that lead to the selection of specific polariton
wavevectors in the planar and transient grating geometry as well as the
Cherenkov angle itself. The formalism can be easily extended to media exhibiting
dispersion in the THz range. Calculations and experimental data for
point-like and planar sources reveal significant differences between the
so-called superluminal and subluminal cases where the group velocity of the
optical pulses is, respectively, above and below the highest phase velocity
in the infrared.
\end{abstract}


\maketitle

\section{Introduction}

Polaritons are hybrid modes of solids and, more generally, dense systems
associated with elementary excitations that carry electromagnetic
polarization.\cite{born54,mills74} Such excitations couple to light,
resulting in a new mixed particle called the polariton. The polariton
description is, in some sense, an alternative to that involving the index of
refraction; polaritons are dressed photons. Phonon polaritons, lattice
vibrations coupled to infrared light, were first studied experimentally by
spontaneous Raman scattering.\cite{henry65} Soon after, mixing of coherent
polaritons with visible laser light was reported for GaP.\cite{faust66} The
generation, dispersion and decay of phonon polaritons have been extensively
investigated using various coherent optical methods, particularly, coherent
anti-Stokes Raman scattering (CARS)\cite{shen65} followed by time-domain CARS%
\cite{gale86} and, more recently, the closely related technique of impulsive
stimulated Raman scattering (ISRS).

This work centers on ISRS, i.e., coherent scattering using pulses whose
bandwidth exceeds the polariton frequency.\cite{yan87a} In ISRS experiments,
a pump pulse imparts an impulsive force on the phonon component of the
polariton starting a coherent oscillation which in turn, through the inverse
process, perturbs the index of refraction.\cite{dhar94,merlin97} This
perturbation is measured by a probe pulse that follows behind the pump pulse
at a controllable time delay. The most common ISRS approach to generate
coherent polariton fields is to use an intensity grating obtained from
crossing two pump beams. This geometry leads to polaritons of nearly
well-defined wavevector even in the pulsed case where the (transient)
grating travels with the pulses.\cite{wiederrecht95,bakker98} An alternative
method, associated with the nonlinear susceptibility $\chi ^{(2)}$, is to
excite the electromagnetic component of the polariton through difference
frequency generation (DFG), or optical rectification.\cite{bass62} Here, the
pump pulse induces a nonlinear polarization proportional to its intensity
envelope. If the pulse is tightly focused and its group velocity is greater
than the phase velocity in the infrared, its polarization leads to
emission of infrared light in much the same way that a relativistic dipole
emits (coherent) Cherenkov radiation (CR). This interpretation, originally proposed for
second-harmonic generation,\cite{tien70} was put forth for DFG by Auston and
coworkers\cite{auston84,kleinman84,auston88} whose technique became later
the standard for generating short THz pulses.\cite{hu90}. Recently, the CR
interpretation was revisited in the analysis of pump-probe experiments on
ZnSe\cite{stevens01} and ZnTe.\cite{wahlstrand02} In these and other polar
materials, strong dispersion near the frequency of the transverse-optical
(TO) phonon results in two qualitatively distinct regimes for CR depending
on whether the velocity of the source is larger or smaller than the phase
velocity of light at zero frequency. In the former, superluminal regime, the
CR field is qualitatively the same as in the experiments of Auston and
co-workers \cite{auston88} whereas the latter subluminal regime displays new
features. \cite{afanasiev99}

In the ultrafast literature, the relationship between the two, transient
grating and tightly-focused single pump methods is rarely discussed. While
the generation mechanisms are closely related, ISRS is often described as
four-wave mixing and, as such, as a technique that distinguishes itself from
DFG. Here, we present a unified approach encompassing the two methods. We
show that the generation of THz radiation by single or multiple pulses of
arbitrary shape can always be described as CR. We also report experimental
results on three materials, ZnTe, GaP, and LiTaO$_{3}$, in three geometries
(point, planar, and transient grating) using an electro-optic field sampling
technique based on that of Auston and co-workers.\cite{auston88} This method
allows us to map out the polariton field as a function of time delay and
lateral position for polarization sources of arbitrary shape and,
simultaneously, to image the shape of the source pulse itself.

\section{Phonon polaritons and optical susceptibilities}

\begin{figure}[tbp]
\centerline{\scalebox{0.98}{\includegraphics{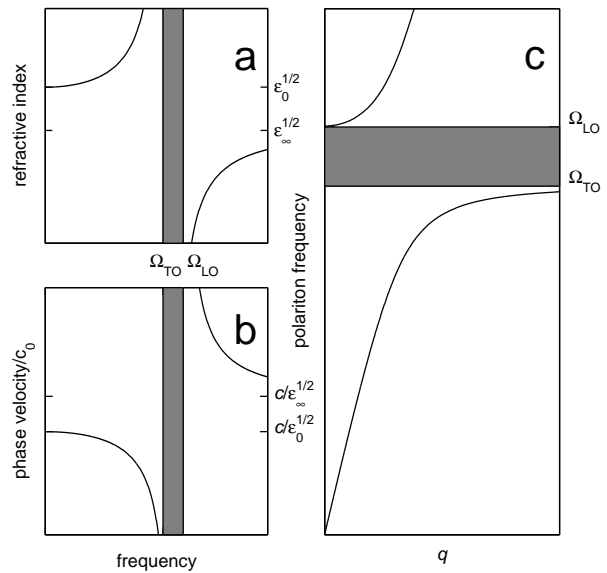}}}
\caption{ Optical parameters for a generic
material with a single triply degenerate infrared-active phonon; see Eq.~(\ref{indexeq}) (a)
Refractive index $n$, (b) phase velocity $c/n$, (c) polariton dispersion.
The darkened regions indicate the band between $\Omega_{\text{TO}}$ and $%
\Omega_{\text{LO}}$ in which light cannot propagate. }
\label{indexfig}
\end{figure}

As mentioned earlier, polaritons can be described either as excitations that
carry polarization through the coupling to the electromagnetic field, or as
light perturbed by the refractive index of the medium, which exhibits strong
dispersion near the resonant TO frequency. In the Cherenkov picture of
phonon polariton generation we take the latter view, incorporating the
effect of the infrared-active phonon into the dielectric function and the
second-order nonlinear susceptibility. For a diatomic cubic lattice, and
neglecting damping, the dielectric function is of the Lorentz form
\begin{equation}
\epsilon (\Omega )=\epsilon _{\infty }+\frac{\epsilon _{0}-\epsilon _{\infty
}}{1-(\Omega /\Omega _{\text{TO}})^{2}},  \label{indexeq}
\end{equation}
where $\Omega _{\text{TO}}$ is the TO frequency of the three-fold degenerate
infrared-active mode, $\epsilon _{\infty }$ is the dielectric constant
including the effects of higher-lying resonances, and $\epsilon _{0}$ is the
static dielectric constant. The Lyddane-Sachs-Teller (LST) formula $\Omega _{%
\text{LO}}^{2}=\frac{\epsilon _{0}}{\epsilon _{\infty }}\Omega _{\text{TO}%
}^{2}$ gives the relationship between $\Omega _{\text{LO}}$, the
longitudinal optical (LO) phonon frequency, and $\Omega _{\text{TO}}$. The
refractive index $n(\Omega )=\sqrt{\epsilon (\Omega )}$, phase velocity $%
c/n(\Omega )$, and polariton dispersion relation $\Omega =cq/n(\Omega )$ are
plotted in Fig.~\ref{indexfig} for a material with $\epsilon (\Omega )$
given by Eq.~(\ref{indexeq}). In this work we deal mainly with frequencies
below $\Omega _{\text{TO}}$. Near zero frequency, where the dispersion curve
is linear, the polariton is mostly light-like whereas, near $\Omega _{\text{%
TO}}$, it is phonon-like. When there is more than one infrared-active
phonon, as in LiTaO$_{3}$, each mode contributes to the polarization and the
dielectric function reflects a sum of these contributions (poles).

Nonlinear interaction of visible light with phonon polaritons involves both
the lattice and the THz electromagnetic field. The nonlinear tensor that
couples polaritons and visible light is thus composed of two parts: the
Raman tensor $\frac{d\chi }{dQ}=\chi ^{\mathrm{R}}$, which couples visible
light to the phonon mode with amplitude $Q$, and the second-order nonlinear
susceptibility $\frac{d\chi }{dE}=\chi ^{(2)}$, which couples visible to
infrared electric fields. Both of these susceptibilities are almost entirely
electronic in nature. Counting both contributions, the total
nonlinear polarization is\cite{faust66} 
\begin{equation}
P_{i}^{\text{NL}}=\chi _{ijk}^{(2)}E_{j}(\Omega )E_{k}(\omega _{L})+\chi
_{ik}^{\text{R}}Q_{j}E_{k}(\omega _{\text{L}})  \label{pol2}
\end{equation}
where $Q_{j}$ is the amplitude of one of the three-fold degenerate phonon
modes, $\omega _{L}$ is the frequency of the visible laser light and $\Omega 
$ is the frequency of the phonon polariton. As is well-known, the Raman
contribution can be included in an effective nonlinear susceptibility $%
d_{ijk}$, each tensor element of which is of the form 
\begin{equation}
d(\Omega )=\chi ^{(2)}[1+C(1-\frac{\Omega ^{2}}{\Omega _{\mathrm{TO}}^{2}}%
)^{-1}],  \label{pol3}
\end{equation}
where the Faust-Henry coefficient $C=\chi ^{\mathrm{R}}(\chi
^{(2)})^{-1}e^{*}\mu ^{-1}(\Omega _{\mathrm{TO}})^{-2}$ gives the ratio of
the Raman (often termed ionic) to the electronic contribution.\cite{faust66}
Here, $e^{*}$ is the transverse effective charge and $\mu $ is the reduced
mass of the lattice mode. Because $\chi ^{\mathrm{R}}$ and $\chi ^{(2)}$ do
not change appreciably with frequency in the THz range, $C$ can be
approximated as a constant for a given laser wavelength and material. If
there is more than one phonon mode, there is a Raman contribution to $d$ and
to the Faust-Henry coefficient from each one. In our experiments we do not
need to consider the dispersion of $d$, for we mainly deal with low
frequency polaritons. Thus, in the following we assume that the nonlinear
susceptibility is constant, and approximately given by $\chi ^{(2)}(1+C)$.

\section{Theory of Cherenkov emission in an infinite crystal}

In non-centrosymmetric media, an optical pulse induces through $\chi ^{(2)}$
a low-frequency polarization proportional to its intensity. This is DFG,
also known as optical rectification.\cite{bass62} The dipolar charge
distribution moves at the optical pulse's group velocity $v_{\text{g}}$ and,
like a charged particle, it can emit CR if $v_{g}$ is greater than the phase
velocity of the THz radiation in the medium. The bandwidth of the THz
radiation emitted by a pulse is limited to the pulse bandwidth. The spatial
and temporal shape of the pulse envelope, together with the optical
properties of the medium, determine the THz field.

Kleinman and Auston adapted the theory of CR to an optical pulse traveling
through a nonlinear isotropic medium.\cite{kleinman84} Their approach, which
we build on here, is essentially to separate the nonlinear mechanism by
which the polarization in the medium is generated from the Cherenkov
mechanism by which infrared light is emitted. This approach works well in
the absence of higher-order effects such as self-focusing and self phase
modulation, which affect the spatiotemporal shape of the pulse as it
propagates, and depletion, which diminishes the induced polarization.
Neglecting these effects, the emission process is entirely analogous to the
emission of CR by a continuum of dipoles traveling at the group velocity
of the pulse. In the following, we discuss electro-optic CR by pulses of a
variety of spatiotemporal shapes.

\subsection{Point source}

For a beam focused to a small waist compared to the polariton wavelength,
the polarization induced by the optical pulse can be approximated by a point
dipole. This geometry, the 3D case discussed by Kleinman and Auston,\cite
{kleinman84} is the closest electro-optic analogue to conventional CR
(emitted by a relativistic monopole), which is coherent. The theory of CR was largely worked
out more than 60 years ago by Tamm and Frank.\cite{frank37,tamm39} For an
infinite medium and pulses propagating along the $z$ axis, the fields are a
function of $z-v_gt$ and the cylindrical coordinates $\rho$ and $\phi$. This
approximation, valid for a thick sample, greatly simplifies the theory.

The radiation field due to a propagating dipole can be derived from the
radiation field due to a monopole by taking a derivative in the direction of
the dipole orientation.\cite{zrelov70} In the experiments discussed here,
the induced dipole is oriented perpendicular to the $z$ axis. For this
orientation, the fields are proportional to $\cos \phi$, where $\phi$ is
measured from the direction of the dipole, dropping for simplicity the $\phi$
dependence, we have\cite{frank37,tamm39,zrelov70}
\begin{eqnarray}
E_z (\rho,z-v_gt) & \propto & \int e^{i\Omega(t-\frac{z}{v_g})}(\frac{c^2}{%
v_g^2 n^2(\Omega)}-1) \frac{\partial a}{\partial \rho} \Omega d\Omega \\
E_\rho (\rho,z-v_gt) & \propto & \int -ie^{i\Omega(t-\frac{z}{v_g})}\frac{1}{%
n^2(\Omega)} \frac{\partial^2 a}{\partial \rho^2} d\Omega,  \label{fieldseq}
\end{eqnarray}
where $E_z$ and $E_\rho$ are the $z$ and $\rho$ components of the electric
field, and 
\begin{equation}
a(\rho,\Omega) = H_0^{(1)}[s(\Omega)\rho]
\end{equation}
where $H^{(1)}_0$ is a Hankel function and $s(\Omega)=\Omega(\epsilon
v_g^2/c^2-1)^{1/2}/v_g=\Omega \tan \theta_C/v_g$.

Radiation is emitted at frequencies for which the Cherenkov condition $%
v_{g}>c/n(\Omega )$ is satisfied. Integrating Eq.~(\ref{fieldseq}) over
these frequencies, two qualitatively different radiation patterns emerge,
depending on $v_{g}$.\cite{stevens02} For $v_{g}>c/n(0)$, the laser pulse
propagates faster than the entire lower polariton branch $0<\Omega <\Omega _{%
\mathrm{TO}}$. We call this the superluminal regime. For $v_{g}<c/n(0)$, the
laser pulse only travels faster for frequencies in the band $\Omega
_{C}<\Omega <\Omega _{\mathrm{TO}}$, where the velocity matched frequency $%
\Omega _{C}$ is defined by $v_{g}=c/n(\Omega _{C})$. We have referred to
this as the subluminal regime.\cite{ginzburg02}

\begin{figure}[tbp]
\centerline{\scalebox{.5}{\includegraphics{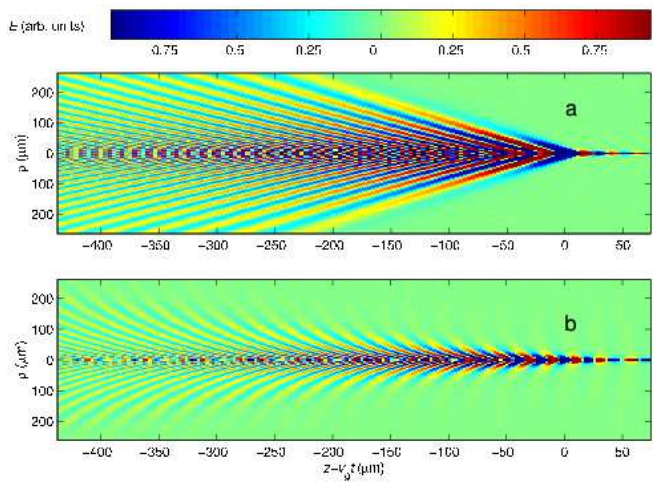}}}
\caption{Calculated $E_\rho$ for a point dipole at $\rho=0$ and $z-v_gt=0$
traveling through a medium with dispersion given by Eq.~(\ref{indexeq}). (a)
Superluminal, $v_g>c/n(0)$, and (b) subluminal, $v_g<c/n(0)$, regimes. }
\label{theory3d}
\end{figure}

Shown in Fig.~\ref{theory3d} are plots of $E_{\rho }$, calculated using Eq.~(%
\ref{fieldseq}), in a medium with refractive index given by Eq.~(\ref
{indexeq}). For the superluminal regime (Fig.~\ref{theory3d}a), there is a
sharp shock front with small amplitude ripples behind it. The weak
ripples are due to dispersion, but the shock wave is defined by $\Omega =0$.
On the other hand, the subluminal regime (Fig.~\ref{theory3d}b) is dominated
by dispersion effects. The angle of the phase front depends on $\rho $, and
there is a large field near the axis of motion $\rho =0$. The radiation
pattern is still confined within a cone, but the cone angle becomes larger as $%
v_{g}$ decreases, in contrast to the superluminal regime. The special
features of the subluminal regime are apparent in the calculations reported
by Afanasiev and co-workers.\cite{afanasiev99}

\begin{figure}[tbp]
\centerline{\scalebox{.98}{\includegraphics{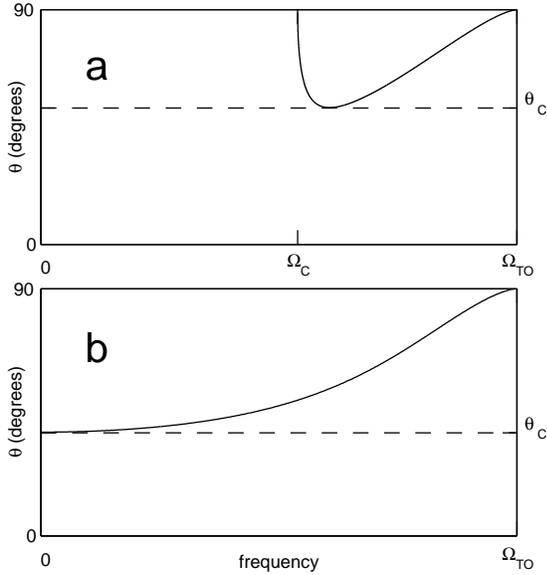}}}
\caption{Plot of angle versus polariton frequency; see Eq.~\ref{angeq}. Dashed
line indicates the Cherenkov angle $\theta_C$ and $\Omega_C$ indicates the
group velocity matched frequency. (a) Subluminal regime. (b) Superluminal
regime. }
\label{angfig}
\end{figure}

In the CR literature, it is common to use the asymptotic form of the fields
for large $\rho $ to find an approximate analytical solution. In this limit, 
$E_{\rho }$ becomes 
\begin{equation}
E_{\rho }\propto \int_{\Omega _{C}}^{\Omega _{\mathrm{TO}}}\frac{i\sqrt{%
2s^{3}(\Omega )}}{n^{2}(\Omega )\sqrt{\pi \rho }}e^{-i[\Omega
(t-z/v_{g})-s(\Omega )\rho +\pi /4]}d\Omega .
\label{eq7}
\end{equation}
For large $\rho $, the integral is small except for values of $\Omega $ for
which the integrand does not oscillate. Using the stationary phase method,
Eq.~\ref{eq7} can be approximated as 
\begin{equation}
E_{\rho }\propto \sum_{\Omega _{i}}\frac{2\sqrt{s^{3}(\Omega _{i})}}{%
n^{2}(\Omega _{i})\rho }\frac{1}{\sqrt{|s^{\prime \prime }(\Omega _{i})|}}%
e^{-i[\Omega _{i}(t-z/v_{g})-s(\Omega _{i})\rho ]},
\end{equation}
where $\Omega _{i}$ are the frequencies within the range of integration for
which the phase in the exponential is stationary. These are the solutions to 
\begin{equation}
v_{g}\frac{ds}{d\Omega }=\frac{v_{g}t-z}{\rho }=\cot \theta ,
\label{rootseq}
\end{equation}
which defines a cone of angle $\theta $ with respect to the $z$ axis. The
cone angle depends on frequency because $s(\Omega )$ contains $n(\Omega )$.
Explicitly, Eq.~(\ref{rootseq}) becomes 
\begin{equation}
v_{g}\frac{ds}{d\Omega }=\Omega ^{2}\frac{v_{g}}{sc^{2}}\frac{d\epsilon }{%
d\Omega }+\frac{v_{g}s}{\Omega }=\cot \theta .  \label{angeq}
\end{equation}
The dependence of $\theta $ on $\Omega $ is plotted for the two regimes in
Fig.~\ref{angfig}. In the superluminal case, $\theta $ is a monotonic
function of frequency, whereas in the subluminal regime two frequencies map
onto the same angle, resulting in the complex beating behavior shown in Fig.~%
\ref{theory3d}b.

The angle $\theta _{C}$, defined as the smallest angle at which radiation is
emitted, is shown as a dashed line in Fig.~\ref{angfig}. The frequencies
associated with this angle are solutions to $s^{\prime \prime }(\Omega )=0$.
In the superluminal regime, $\epsilon (\Omega )v_{g}^{2}/c^{2}-1>0$ for all
frequencies within the range of integration, and $\Omega =0$ is the only
solution. Thus, from Eq.~(\ref{angeq}), $\cos \theta _{C}=v_{g}n(0)/c$,
which is the standard Cherenkov formula. In the subluminal regime, there are
two solutions. For a material with dispersion given by Eq.~(\ref{indexeq}), $%
\theta _{C}=\tan ^{-1}(4\gamma ^{-1}\sqrt{(\eta -\zeta )/\zeta ^{3}})$,
where $\gamma =(1-\epsilon _{\infty }v_{g}^{2}/c^{2})^{-1/2}$, $\zeta
=2-(4-3\eta )^{1/2}$, and $\eta =1-(\Omega _{C}/\Omega _{\mathrm{TO}})^{2}$.%
\cite{afanasiev99}

So far, we have considered isotropic materials, and therefore, the
considerations above apply to ZnTe and GaP but, in the strictest sense, not
to the uniaxial LiTaO$_{3}$. The theory of CR in an anisotropic material is
considerably more complex than that discussed here, because extraordinary
and ordinary waves may be emitted. Other complications result for the case of an
extremely tightly focused pulse, for which there exists a small
longitudinal component of the laser electric field, leading to the breakdown
of the point dipole approximation. In addition, the confocal parameter comes
into play resulting in the breakdown of the infinite crystal approximation.
We note in passing that exact integrals for an electric dipole oriented
arbitrarily with respect to the axis of motion, as well as those of a
magnetic dipole, are given in the work of Zrelov.\cite{zrelov70}

\subsection{Planar source}

When the beam waist $w$ is much larger than the wavelength of the polariton (%
$w>>2\pi c n/\Omega$), we have another important limiting case. Here, the
polarization induced by the optical pulse is shaped like a pancake. This
planar geometry is often used for the generation and detection of THz pulses
by nonresonant optical rectification.\cite{wu96,nahata96} When applied to
this problem, the CR picture is useful because it quantifies phase matching
issues encountered in thick crystals due to dispersion in the infrared.

For an infinitely thick crystal, and approximating the optical pulse as
infinitely short (this is valid as long as the pulse is short compared to
the period of the radiation being generated) the polarization induced is $%
\mathbf{P}\approx \zeta \delta (z-v_{g}t)\mathbf{e}_{d}$, where $\mathbf{e}%
_{d}$ is a unit vector pointing in the direction of the dipole orientation
and $\zeta $ is the areal polarization. It is possible to obtain the CR
field from the point dipole solution by convolution, but here we use the
corresponding Hertz potential 
\begin{equation}
\mathbf{\Pi }(\Omega )=-\frac{4\pi \zeta e^{-i\Omega (t-z/v_{g})}}{%
v_{g}\epsilon (\Omega )s^{2}(\Omega )}\mathbf{e}_{d}.
\end{equation}
The electric field is then\cite{stevens01} 
\begin{eqnarray}
\mathbf{E} &=&\int_{0}^{\Omega _{\mathrm{TO}}}\frac{\epsilon (\Omega )\Omega
^{2}}{c^{2}}\mathbf{\Pi }(\Omega )d\Omega  \\
&=&-\frac{4\pi \zeta }{v_{g}c^{2}}\mathbf{e}_{d}\int_{0}^{\Omega _{\mathrm{TO%
}}}\frac{\Omega ^{2}e^{-i\Omega (t-z/v_{g})}}{s^{2}(\Omega )}d\Omega .
\end{eqnarray}
It can be shown that the field vanishes unless $s^{2}(\Omega )=0$. In the
superluminal regime, $s^{2}(\Omega )>0$ for all $\Omega $, and $\mathbf{E}=0$%
. In the subluminal regime, $s^{2}(\Omega _{C})=0$. Note that the phase
matching condition is \emph{only} met in the presence of dispersion. In the
superluminal regime, no radiation is emitted from the bulk because the pulse
is traveling faster than the phase velocities of all infrared frequencies.

\begin{figure}[tbp]
\centerline{\scalebox{.5}{\includegraphics{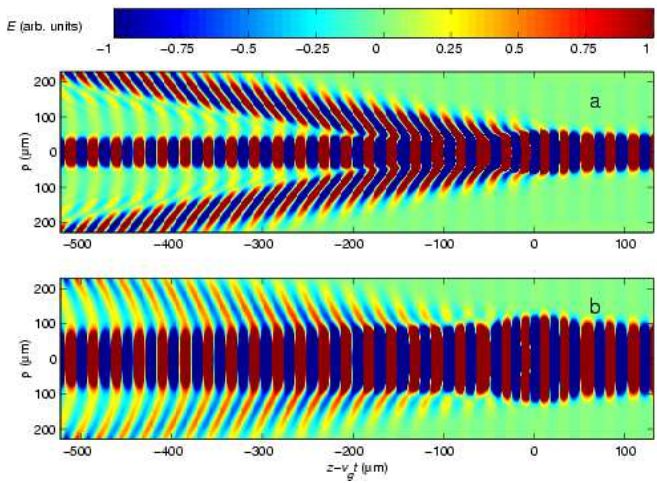}}}
\caption{Calculated $E_\rho$ for an extended source of width $w$,
obtained by convoluting results for a point dipole with a Gaussian in the $\rho$
direction (subluminal regime).
(a) $w=20$ $\mu$m,
(b) $w=40$ $\mu$m
}
\label{pltheoryfig}
\end{figure}

Practically, a beam cannot be focused to a point, nor is it really planar.
If cylindrical lenses are used to focus the beam, it can be extended in one
direction and tightly focused in another. The extreme case of
the 2D geometry was treated by Kleinman and Auston.\cite{kleinman84} The point
geometry solution is the Green's function of the problem, in that the
radiation pattern of an arbitrary source can be obtained by convolution of the
point solution with the source shape. Two examples are shown in Fig.~\ref{pltheoryfig}. This is a powerful machinery
for calculating the radiation field generated by ultrafast optical pulses
which have been shaped spatially and temporally, for example, in the manner
recently demonstrated by Koehl, Adachi, and Nelson.\cite{koehl99} To the
best of our knowledge, our approach has not been previously applied to this
problem.

\subsection{Periodic source (transient grating)}

\begin{figure}[tbp]
\centerline{\scalebox{.5}{\includegraphics{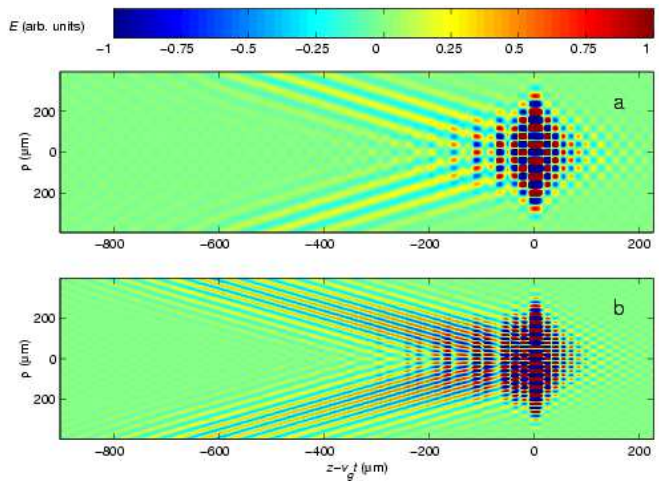}}}
\caption{Calculated $E_\rho$ due to a transient grating, showing the
checkerboard polariton field pattern generated behind the grating.
The grating wavevector in (a) is smaller than in (b).
}
\label{checkfig}
\end{figure}

The last source shape we consider is a periodic distribution, or transient
grating, implemented by interfering two pump beams inside the sample. This
geometry is useful because it excites a unique polariton wavevector
determined by the grating period. If we cross two pump beams of equal
intensity at an angle $\alpha$, Gaussian in space with waists $w_{x}$ in the 
$x$ direction and $w_{y}$ in the $y$ direction and Gaussian in time with
pulsewidth $\tau $, the intensity grating has the form 
\begin{equation}
I(x,y,z-v_{g}t)\propto e^{-x^{2}/w_{x}^{2}+y^{2}/w_{y}^{2}}\cos (\frac{k_{L}x%
}{\sin \alpha })e^{-(z/v_{g}-t)^{2}/\tau ^{2}}
\end{equation}
where $k_{L}$ is the wavevector of the laser light inside the medium. Here,
we have assumed that $\alpha$ is small, which makes $v_{g}$ approximately equal to the
group velocity of each separate pulse. This is a good approximation for the
experiments described here, since $k_{L}$ is much larger than the wavevector
of the gratings. The Cherenkov angle converts the spatial periodicity of the
transient grating into a temporal periodicity in the polariton field,
resulting in a checkerboard pattern and a well-defined polariton frequency $%
\Omega$, as shown in Fig.~\ref{checkfig}.

In the impulsive limit, a defocused probe pulse traveling at the same speed
as the pump pulse and, thus, at the speed of the polariton field, samples a
slice of the checkerboard field in the $xy$ plane. The electric field in
this slice is periodic with the same wavevector $k_{x}=k_{L}/\sin \alpha $ as
the intensity grating. Through the electro-optic effect, the probe pulse
diffracts from the periodically varying index of refraction, resulting in an
oscillatory signal at $2\Omega $ as a function of time delay, because the
diffracted intensity is proportional to $|E|^{2}$. This technique is known
as homodyne detection in the ISRS literature.\cite{wiederrecht95} Heterodyne
detection,\cite{bakker98,gautier96,crimmins02} in which one mixes the
diffracted signal with undiffracted probe light, gives a much larger signal
at $\Omega $, because the signal is proportional to $E$. Because it is
sensitive to the phase of the polariton field, which defines the phase of
the diffracted probe pulse, heterodyne detection works just as well outside
the checkerboard, in what we might call the Cherenkov wing of the field.
As we discuss in detail below, heterodyne detection is often discussed in terms of four-wave mixing.\cite
{gautier96}

It is important to emphasize that the wavevector of the grating is not the
same as the polariton wavevector, since the polariton propagates in the
direction normal to the wavefront. The magnitude of the polariton wavevector
is given by $q=k_{x}\sin \theta $, where $\theta $ is the Cherenkov angle.
In many early ISRS papers, it is assumed that the wavevector of the grating
is the same as the polariton wavevector, tantamount to assuming that $\theta 
$ is very large. This is generally a poor approximation, and leads to
incorrect polariton dispersion measurements, as recently pointed out by
Hebling.\cite{hebling02}

The simple interpretation above works best for large $w$, in which case the
polariton wavevector is well-defined. For small $w_{x}$, edge effects make
the polariton field more complex but, by convolving the point dipole
solution with the grating (as in Fig.~\ref{checkfig}), we can calculate the
polariton field due to a grating of arbitrary shape.

\section{Phase matching}

One source of confusion when comparing the transient grating, the tightly
focused, and the planar geometries is that, at first glance, the phase
matching conditions seem entirely different. Here, we clarify this point. An
ultrafast pulse contains a continuum of frequencies and wavevectors, and the
generated polariton wavevectors are determined by conservation of energy and momentum. The Cherenkov
interpretation is, of course, consistent with phase matching. In all geometries, the
nonlinear polarization contains terms of the form $e^{i(\mathbf{k}_{2}-%
\mathbf{k}_{1}).\mathbf{r}-i(\omega _{2}-\omega _{1})t}$, where $\mathbf{k}%
_{1}$, $\mathbf{k}_{2}$, $\omega _{1}$, and $\omega _{2}$ denote wavevector
and frequency components of the optical pulse. For the planar geometry, the
wavevector components all point along the $z$ axis, and conservation of
momentum requires $q=k_{2}-k_{1}=[\omega _{2}n(\omega _{2})-\omega
_{1}n(\omega _{1})]/c$, where $q$ is the polariton wavevector. Since $\Omega
=\omega _{1}-\omega _{2}$, 
\begin{eqnarray}
q &=&\frac{\omega _{2}n(\omega _{2})-(\omega _{2}+\Omega )n(\omega
_{2}+\Omega )}{c} \\
&\approx &\frac{\Omega }{c}[n(\omega _{2})+\omega _{2}n^{\prime }(\omega
_{2})]=\frac{\Omega }{v_{g}(\omega _{2})},
\end{eqnarray}
making the connection to the group velocity matching argument derived in
Section IIIB.

A tightly focused optical pulse has wavevector components pointing in many
directions. The dependence of the polarization on $z-v_gt$ forces $q_z =
\Omega/v_g$. The polariton wavevector and frequency must also satisfy $%
\Omega=|\mathbf{q}|c/n(\Omega)$. Let $q_\rho$ be the component of the
polariton wavevector perpendicular to the $z$ axis. Momentum conservation
gives 
\begin{equation}
\frac{q_\rho}{q_z}=\frac{\sqrt{q^2-q^2_z}}{\Omega/v_g}=\sqrt{\frac{%
n^2(\Omega) v_g^2}{c^2}-1},  \label{pmeq}
\end{equation}
and the polariton is emitted at an angle $\tan^{-1} q_\rho/q_z$ to the $z$
axis, which is identical to the Cherenkov angle.

In the transient grating geometry, the angle between $\mathbf{k}_{1}$ and $%
\mathbf{k}_{2}$ is well-defined, and the wavevector $\mathbf{q}=\mathbf{k}%
_{1}-\mathbf{k}_{2}$ is emitted in the direction given by Eq.~(\ref{pmeq})
because the polarization is a function of $z-v_{g}t$. The transient grating
configuration allows one to excite wavevectors that are higher than those
accessible in the planar geometry.

A recently proposed technique developed by Hebling and co-workers relies on
a tilted pulse front.\cite{hebling02a} This can also be encompassed in a
Cherenkov radiation framework as a planar source which propagates at an
angle. Tuning the tilt angle changes the wavevector of the generated
polariton, and this shows promise as an easily tunable THz source.

As is well-known, it is possible to interpret ISRS as four wave mixing.\cite
{gautier96} Here one hides both the generation and detection of the
polariton in $\chi^{(3)}_{\mathrm{eff}}$, much as we included the Raman
contribution to the generation of the polariton in $\chi^{(2)}$. The
four-wave mixing polarization is 
\begin{eqnarray}
P(\omega_s) & = & \chi^{(3)}_{\mathrm{eff}}(\omega_p,\Omega-\omega_p,%
\omega_p) E_1(\omega_1) E^*_2(\omega_1-\Omega) E_p(\omega_p)  \nonumber \\
& = & d(\Omega)[d(\Omega) E_1(\omega_1) E^*_2(\omega_1-\Omega)] E_p(\omega_p)
\end{eqnarray}
where $E_1$ and $E_2$ are the pump fields, $E_p$ is the probe field, and $d$
is the effective nonlinear susceptibility defined in Section II. All
propagation and damping effects can be included in $\chi^{(3)}_{\mathrm{eff}}
$, and this formalism is often used.\cite{gale86}

Finally, we discuss briefly the advantages and disadvantages of the various
scattering geometries. The experiments with tightly focused beams produce
the highest peak polariton amplitude (for a given pulse peak power), but
also a wide range of wavevectors and frequencies. ISRS by a transient
grating has the advantage of defining the polariton wavevector and producing
a large amplitude polariton at that particular wavevector. CARS has the
advantage of yielding the best frequency and wavevector resolution.

\section{Materials}

Generation of electro-optic CR requires a non-zero $\chi^{(2)}$, present
only in crystals without inversion symmetry. In our experiments, we used
three readily available and well-characterized materials: the zincblende
ZnTe and GaP, and LiTaO$_3$. The parameters relevant to our work are listed
in Table \ref{mattab}.

\begin{table}
\begin{center}
\begin{tabular}{c c c c} \hline
Parameter     & GaP  & ZnTe & LiTaO$_3$\footnote{Parameters are for the lowest $A_1$ mode.} \\ \hline \hline
$\Omega_{\rm TO}/2\pi$(THz) & 11.0\footnotemark\footnotetext[2]{Reference \onlinecite{mooradian66}.} & 5.32\footnotemark\footnotetext[3]{Reference \onlinecite{gallot99}.} & 6\footnotemark\footnotetext[4]{Reference \onlinecite{barker70}.} \\
$\Omega_{\rm LO}/2\pi$(THz) & 12.1\footnotemark[2] & 6.18\footnotemark[3] & 12\footnotemark\footnotetext[5]{From LST relation using parameters of Ref. \onlinecite{barker70}.} \\
$\sqrt{\epsilon_0}$ & 3.31\footnotemark\footnotetext[6]{Reference \onlinecite{pikhtin76}.} & 3.16\footnotemark[3] & 6.3\footnotemark[4] \\
$n_g$ at 800 nm & 3.56\footnotemark[6] & 3.24\footnotemark\footnotetext[7]{Reference \onlinecite{marple64}.} & 2.2 \\
$\Omega_C/2\pi$\footnotemark\footnotetext[8]{Calculated from $\epsilon_\infty$.}(THz) & 7.7  & 2.2 & - \\
\hline
\end{tabular}
\end{center}
\caption{Table of parameters used in the calculations.}
\label{mattab}
\end{table}

\subsection{GaP and ZnTe}

Materials with the zincblende structure have a triply degenerate
infrared-active optical mode. For light normally incident on a (110) face,
the selection rules forbid excitation of the nondispersive LO mode. For the
TO mode, 
\begin{equation}
R_{1}=\left( 
\begin{array}{ccc}
0 & 0 & -b \\ 
0 & 0 & b \\ 
-b & b & 0
\end{array}
\right) \text{ and }R_{2}=\left( 
\begin{array}{ccc}
0 & b & 0 \\ 
b & 0 & 0 \\ 
0 & 0 & 0
\end{array}
\right) ,
\end{equation}
where $R_{1}$ and $R_{2}$ are the Raman tensors for the modes polarized
along the [$1\bar{1}0$] and [001] directions, respectively. The single
independent tensor component $b$ corresponds to the nonlinear coefficient $%
d_{41}$. The dipolar distribution induced by the pump pulse is always
oriented perpendicular to the [110] direction. We measure the polariton by
the inverse process, and we are only sensitive to the electric field
component $E_{\rho }$.

In their pioneering work using nonlinear mixing of visible and infrared cw
laser light, Faust and Henry measured $C=-0.47$ in GaP.\cite{faust66} This
leads to total destructive interference between the ionic and electronic
contributions to $d$ at 6.5 THz. Because this frequency is close to $\Omega
_{C}$ for Ti:sapphire pulses, this results in a small signal. Recently
Leitenstorfer and co-workers found $C=-0.07$ in ZnTe by analyzing data from
time-domain spectroscopy of THz pulses by electro-optic sampling.\cite
{leitenstorfer99} This small value leads to $d=0$ at 5.2 THz, very close to $%
\Omega _{\mathrm{TO}}$. For our experiment in ZnTe, we are only able to
measure up to roughly 4 THz, not high enough to probe effects of the
dispersion of $d$.

\subsection{LiTaO$_3$}

This material crystallizes in the perovskite structure. LiTaO$_3$ is
ferroelectric at room temperature and exhibits 4 $A_1$ symmetry and 9 $E$
symmetry modes, all of which are Raman and infrared active.\cite{Raptis88}
There has been considerable disagreement over the assignment of several
Raman peaks, but the lowest-frequency $A_1$ mode considered here is
well-characterized. The Raman tensor for $A_1$ phonons is\cite{Raptis88} 
\begin{equation}
A_1= \left( 
\begin{array}{ccc}
a & 0 & 0 \\ 
0 & a & 0 \\ 
0 & 0 & b
\end{array}
\right).
\end{equation}
The independent tensor elements $a$ and $b$ are associated with nonlinear
coefficients $d_{31}$ and $d_{33}$, respectively. Here, we use $%
\chi^{(2)}=d_{33}$ and $\chi^{\mathrm{R}}=b$, which are the largest
coefficients, to both generate and detect the polariton. The dipolar
distribution points along the optic axis, and the probe pulse is sensitive
to the component of the Cherenkov electric field along that axis. Barker,
Ballman, and Ditzenberger measured the infrared reflectivity and found that,
for the extraordinary index, most of the oscillator strength in the infrared
is in the lowest-lying $A_1$ TO mode.\cite{barker70} This mode is also the
strongest Raman scatterer. We performed our experiment on the lowest
polariton branch.

The refractive index in the visible is weakly anisotropic. Since the laser
pulses in our experiment are polarized along the optic axis, there is no
effect of this anisotropy, in contrast with experiments on $E$ modes.\cite
{albert96} The static dielectric constant is also weakly anisotropic.
However, since the lowest-lying $A_{1}$ and $E$ TO frequencies are different
(6 and 4.2 THz, respectively), the dielectric function becomes more
anisotropic at large frequencies, where the cubic approximation is expected
to break down. For the low frequencies discussed here, the approximation is
well obeyed. For the lowest-lying $A_{1}$ mode in LiTaO$_{3}$, $C$ is
positive.\cite{boyd73}

\section{Experiments}

\subsection{Techniques}

\begin{figure*}[tbp]
\centerline{\scalebox{.7}{\rotatebox{270}{\includegraphics{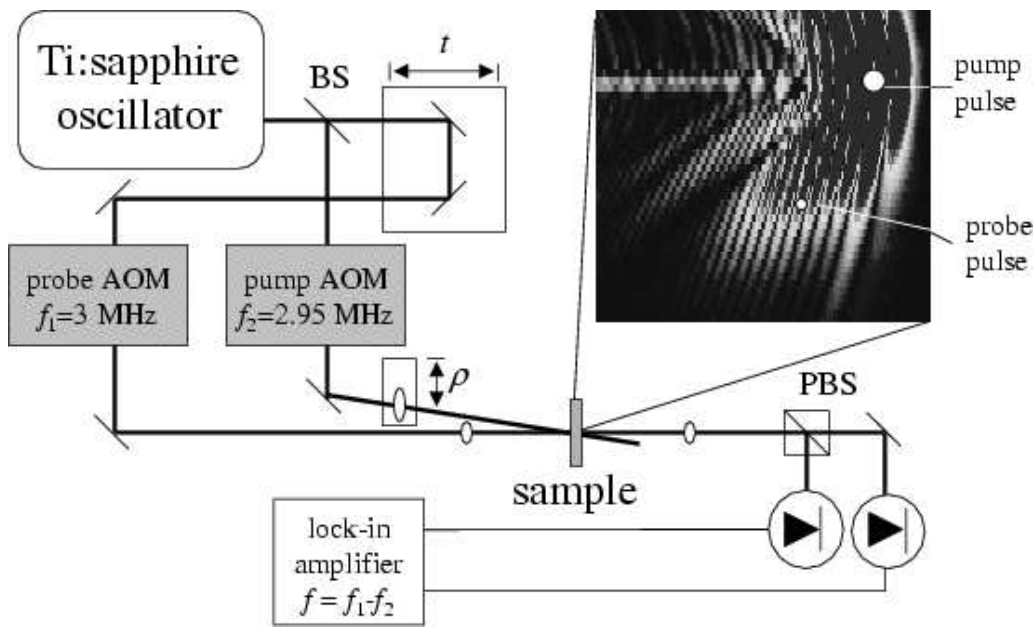}}}}
\caption{ Experimental setup. Pulses split into pump and probe beams,
the time delay $t$ between the pulses is varied with an automated
translation stage. The relative position of the two beams $\rho$ is adjusted
by moving a single lens. The inset shows a close-up of the geometry inside the crystal.
The pump and probe pulses travel along parallel paths inside the crystal at the
same group velocity $v_g$. BS and PBS denote, respectively, a 10:1 beam
splitter and a polarizing beam splitter. Not shown are $\lambda/2$ waveplates, used to
rotate the polarization of the pump and probe beams, and a $\lambda/4$ waveplate
in the probe path which resolves the induced birefringence due to the electro-optic
effect.}
\label{exptsetupfig}
\end{figure*}

Our experimental technique is similar to that of Auston and Nuss.\cite{auston88}
A diagram of the experimental setup is shown in Fig.~\ref{exptsetupfig}.
Pump and probe beams are focused and crossed inside the
sample. The pump pulse is the Cherenkov source and the tightly focused probe
serves as the detector of the Cherenkov radiation field. The probe pulse
``surfs'' on top of the Cherenkov wakefield generated by the pump pulse and
samples the polariton field through both the linear electro-optic and the Raman effects, which
both cause changes in the refractive index. We change the time delay ($t$)
between pump and probe with a motorized delay stage in the probe path and
the relative focal position ($\rho $) between pump and probe by moving a
motorized translation stage upon which the pump's focusing lens is mounted.
By sampling the diffracted probe beam as a function of $t$ and $\rho $, we
map the polariton field.

We used two different detection schemes. The first is a frequency shift
measurement, which takes advantage of the fact that the spectrum of the
probe is shifted proportional to $dn/dt$.\cite{dhar94} The polariton induces
a change in the index of refraction due to its electromagnetic and vibrational components,
so that $dn/dt$ contains a term proportional to $dE/dt$ and another one to $dQ/dt$.
We measure the frequency shift by cutting
half of the probe spectrum with a bandpass filter before detection with a
photodiode. This technique is especially useful when the pump and probe
polarizations are chosen perpendicular to each other to suppress the
scattered pump light. The second technique, widely used for detecting THz
radiation, is polarization detection.\cite{valdmanis82,wu96} When properly
oriented, the polariton's electric field induces birefringence, which can be
probed by measuring the polarization state of the probe beam after the
sample. This yields a signal directly proportional to the THz Cherenkov
electric field.

The spot size inside the crystal was determined experimentally using the
same technique. In ZnTe and GaP, two photon absorption depletes the probe
when pump and probe overlap temporally and spatially, creating a large
Gaussian peak in the signal. In LiTaO$_3$, the intensity-dependent
refractive index due to $\chi^{(3)}$ modulates the probe pulse's phase when
it overlaps with the pump pulse, also producing a large signal (known as the
coherent artifact).

Auston and Nuss used thin samples and a collinear geometry to achieve
spatial resolution on the order of 1.5 $\mu$m.\cite{auston88} A problem
with this setup is that it is difficult to measure the field behind the
source, because of strong pump scattering due to the fact that the two beams
are collinear. In order to measure the source itself plus the field behind
it, we used the nearly-collinear setup described above, which enables the
spatial separation of the pump and probe beams even when they overlap
spatially inside the sample. This introduces a source of distortion in the
data due to spatial walk-off between the pump and probe pulses as they
travel through the sample. To minimize this walk-off, we focused the pump
and probe beams with two small lenses placed as close together as possible.
In order to ensure that the walk-off did not affect the spatial resolution,
we made the plane shared by the pump and probe beams perpendicular to the
axis along which the pump lens was moved. We accomplished this by directing
the probe beam below the pump lens, and moving the relative focal position
horizontally. The resolution in $z-vt$ is determined by the pulsewidth of
the probe pulse, and in $\rho $ by the focal waist of the probe beam.

Ultrafast pulses were generated by a Ti:sapphire oscillator producing 800
nm, 60 fs pulses at the repetition rate of 82 MHz. Acousto-optic amplitude
modulators were used to modulate the pump and probe beams at 3 MHz and 2.95
MHz, respectively, and the probe was detected with an unbiased photodiode at
the difference frequency, 50 kHz, with a lock-in amplifier. The maximum pump
and probe average power at the sample were $\sim$140 and 30 mW,
respectively, limited mostly by the efficiency of the modulators. We
emphasize the importance of the double-modulation technique because it helps
reject the scattered pump light. Recently, experiments have been performed
using a Zernike method with a CCD camera, to image the phase change of a
weakly-focused probe beam caused by the polariton field as a function of
position and time delay.\cite{koehl99,koehl01a} This is another powerful
technique for imaging the Cherenkov field, although it suffers from unwanted
large scattering near the pump pulse.

\subsection{Point source: Experiments}

If the pump pulse is focused to a waist smaller than the wavelength of the
polariton, we can approximate it as a point dipole. Because of the large
thickness of our samples, we are limited to a focal waist of about 20 $\mu $%
m which corresponds to a maximum polariton frequency of 2-4 THz, depending
on the material. For lower frequencies, the pump pulse acts like a
point dipole.

\begin{figure}[tbp]
\centerline{\scalebox{.5}{\includegraphics{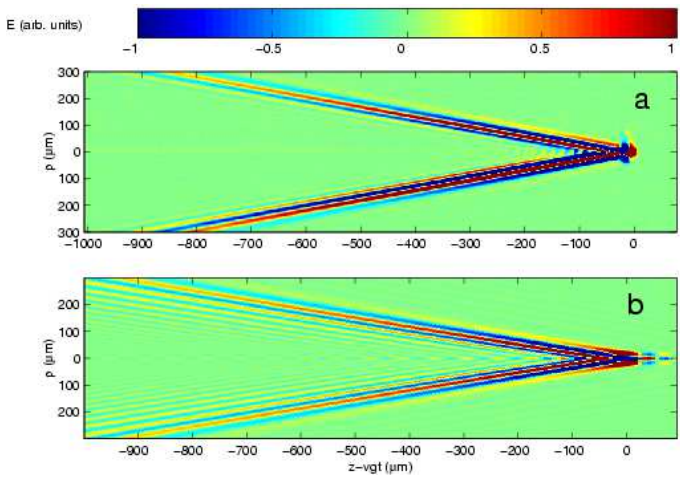}}}
\caption{(a) Experimental results for LiTaO$_3$. The pump waist is 18 $\mu$%
m. (b) Convolution of the point dipole field with a 18 $\mu$m Gaussian. }
\label{ptlitafig}
\end{figure}

First we discuss the data on LiTaO$_{3}$, a superluminal material; see Fig.~%
\ref{ptlitafig}. Our sample was an $x$-cut wafer 2 inches in diameter and 1
mm thick, and we used frequency shift detection. For pulses of wavelength
800 nm, $\theta _{C}=69^{\circ }$. The experiment matches the calculations
quite well after taking into account the 20 $\mu $m pump waist by
convoluting the point dipole solution with a Gaussian of that width.
However, the absence of ripples behind the sharp Cherenkov cone indicates that damping, not included in the simulation, is strong
for higher frequencies. In addition to polaritons of the lowest branch, we were also
able to generate and detect polaritons and coherent phonons from higher
branches; see later.

\begin{figure}[tbp]
\centerline{\scalebox{.5}{\includegraphics{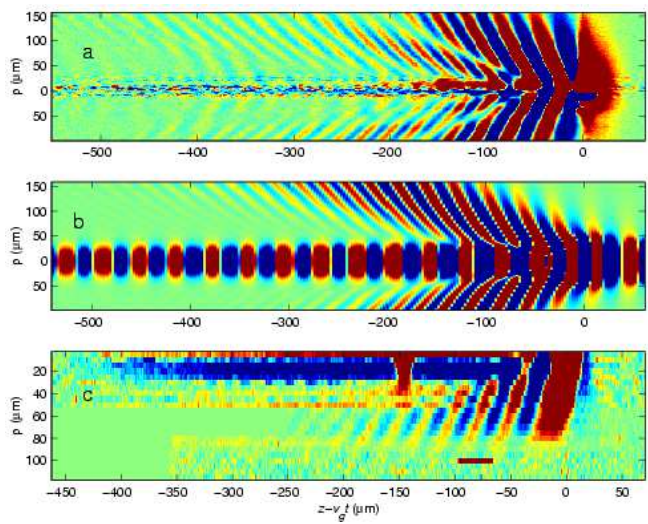}}}
\caption{ (a) Results of a pump-probe differential transmission experiment
in ZnTe. An overall constant background was subtracted for each time scan,
and a decaying exponential was subtracted out for time scans near $\rho=0$.
The feature 1.6 ps behind the pump pulse is an artifact. (b) Calculation
for a point dipole convoluted with a 22 $\mu$m Gaussian in the transverse
direction. (c) Experimental data for a 300 $\mu$m thick ZnTe
sample. }
\label{ptzntefig}
\end{figure}

Our measurements on ZnTe, a subluminal material, and the comparison with
calculations are shown in Fig.~\ref{ptzntefig}. The (110) oriented ZnTe
sample was 10$\times$10$\times$1 mm$^3$, and polarization-sensitive
detection was used. The data shows many of the features predicted for
subluminal CR. The artifact that appears about 1.6 ps after the pump pulse
is due to a second pulse about 1000 times smaller than the pump pulse, due
to a reflection from an optical element upstream. We have subracted the
contribution from an exponentially decaying signal caused by free carriers
generated by two-photon absorption. This signal is partially balanced out by
using polarization-sensitive detection, and the remnant has been fitted and
subtracted out in Fig.~\ref{ptzntefig}a to show the absence of CR near $\rho
=0$.

As discussed earlier, subluminal CR is expected to exhibit a beating
pattern. In Fig.~\ref{ptzntefig}a, there is a striking node that separates
the shock wave into two and possibly three distinct parts. This can be
explained in the calculations by both subluminal beats, or by a convolution
artifact due to the finite spatial size of the pump and probe pulses. However,
note that the node is not as prominent in the calculation.
This discrepancy, and the lack of experimental signal near $\rho =0$, is
probably due to the finite size of the crystal, as discussed in detail later.

\subsection{Planar source: Experiments}

\begin{figure}[tbp]
\centerline{\scalebox{.9}{\includegraphics{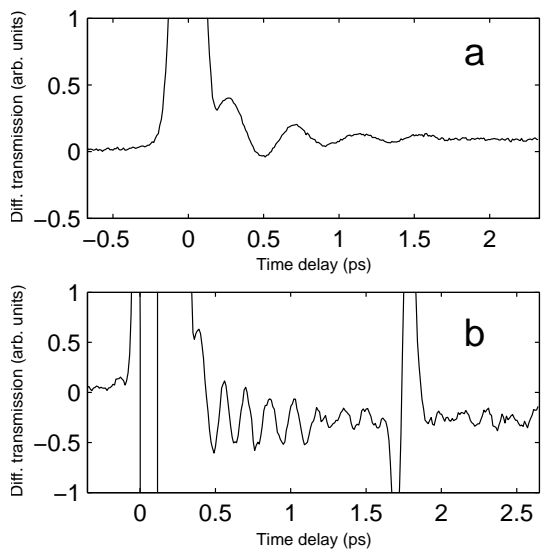}}}
\caption{Results of pump-probe experiments with weakly focused ultrafast
pulses in (a) ZnTe and (b) GaP. The region near time zero is dominated by two photon
absorption. (a) 1 mm thick ZnTe crystal. The main frequency is roughly 2
THz. (b) 1 mm thick GaP crystal. The main frequency is roughly 7.5 THz. The
feature at 2.2 ps is an artifact due to a reflected pulse. }
\label{finite1dfig}
\end{figure}

For the planar geometry, scans were performed only along the $\rho $ axis in
the $z-v_{g}t$ direction. Results for GaP and ZnTe with a pump waist of
roughly 100 $\mu $m are shown in Fig.~\ref{finite1dfig}. The GaP sample,
(110) oriented, was a 4$\times $1$\times $1 mm$^{3}$ crystal. The time scans
show oscillations at the group velocity matched frequencies predicted by
theory, with an apparent decay probably due to a combination of the finite
length of the sample (see below), and the finite extent of the pump pulse.%
\cite{stevens01} The planar geometry is commonly used for generating THz
pulses by nonresonant optical rectification.\cite{nahata96} The distortion
of the electro-optic sampling signal due to infrared dispersion, especially
in a thick crystal, has been the subject of much discussion recently.\cite
{bakker98a,gallot99a}

We also performed an experiment on the superluminal material LiTaO$_{3}$. As
expected, there was no signal from the lowest polariton branch. However, our
pulses were short enough to see higher branches. The phase matching
condition is given by $\Omega /v_{g}$, and this expression was found to
match the higher polariton branches quite well.

\subsection{Finite crystal effects}

Point geometry experiments were also performed on a 300 $\mu $m thick ZnTe
sample. As is apparent in Fig.~\ref{ptzntefig}c, the node we observed in the
1-mm-thick crystal has now disappeared. This is probably a result of the
breakdown of the infinite crystal approximation. In the superluminal regime,
the infinite crystal calculation works well because the radiation is emitted
at large angles at all frequencies down to zero. Therefore, effects of the
finiteness of the crystal can be expected only at the very edge of the shock
wave, at $\rho =L\tan \theta _{C}=Ln_{0}/c$, where $L$ is the length of the
crystal. For example, the edge of the shock wave is at $\rho \approx 2.6$ mm
in our 1 mm thick LiTaO$_{3}$ sample, well beyond what we can practically
measure due to the fall-off proportional to $\rho ^{-1/2}$ in the electric
field from a point source.

The situation is considerably different in the subluminal regime because
polaritons near $\Omega _{C}$ are emitted in the near forward direction. In the
1 mm thick ZnTe crystal, the Cherenkov angle for 2.5 THz polaritons is about 
$9^{\circ }$, so that this component reaches $\rho =160$ $\mu $m before the
pulse reaches the end of the crystal. In the 300 $\mu $m thick crystal, it
only reaches $\rho =50$ $\mu $m. Thus, the subluminal beats, due to
interference between frequency components sent in the forward direction,
require some propagation time to develop. We believe this is the reason why
they do not appear in the thin crystal. This could also explain other
discrepancies between theory and experiment in Fig.~\ref{ptzntefig}. The
large oscillatory signal at $\Omega _{C}$ near $\rho =0$ in the calculation
is clearly absent in the experiment, probably because radiation at $\Omega _{C}$ is emitted
in the forward direction The difference in the shape of the cone between
theory and experiment is likely due to the same effect. A finite crystal
calculation is probably needed to fully understand the details of the
experiment. Note that CR in a finite crystal is similar to the so-called
``Tamm problem'' of CR for a particle traveling along a finite path.\cite
{tamm39,afanasiev00}

The finiteness of the crystal is expected to play an even more important
role in the planar geometry, in which Cherenkov components are emitted
exclusively in the forward direction. The CR pattern is the result of
interference among these components, so that this pattern should change
dramatically with crystal thickness. A very thin crystal would show no
propagation effects, simply reproducing the time derivative of the pump
pulse envelope as a short THz pulse.

\subsection{Periodic source: Experiments}

\begin{figure*}[tbp]
\centerline{\scalebox{.9}{\includegraphics{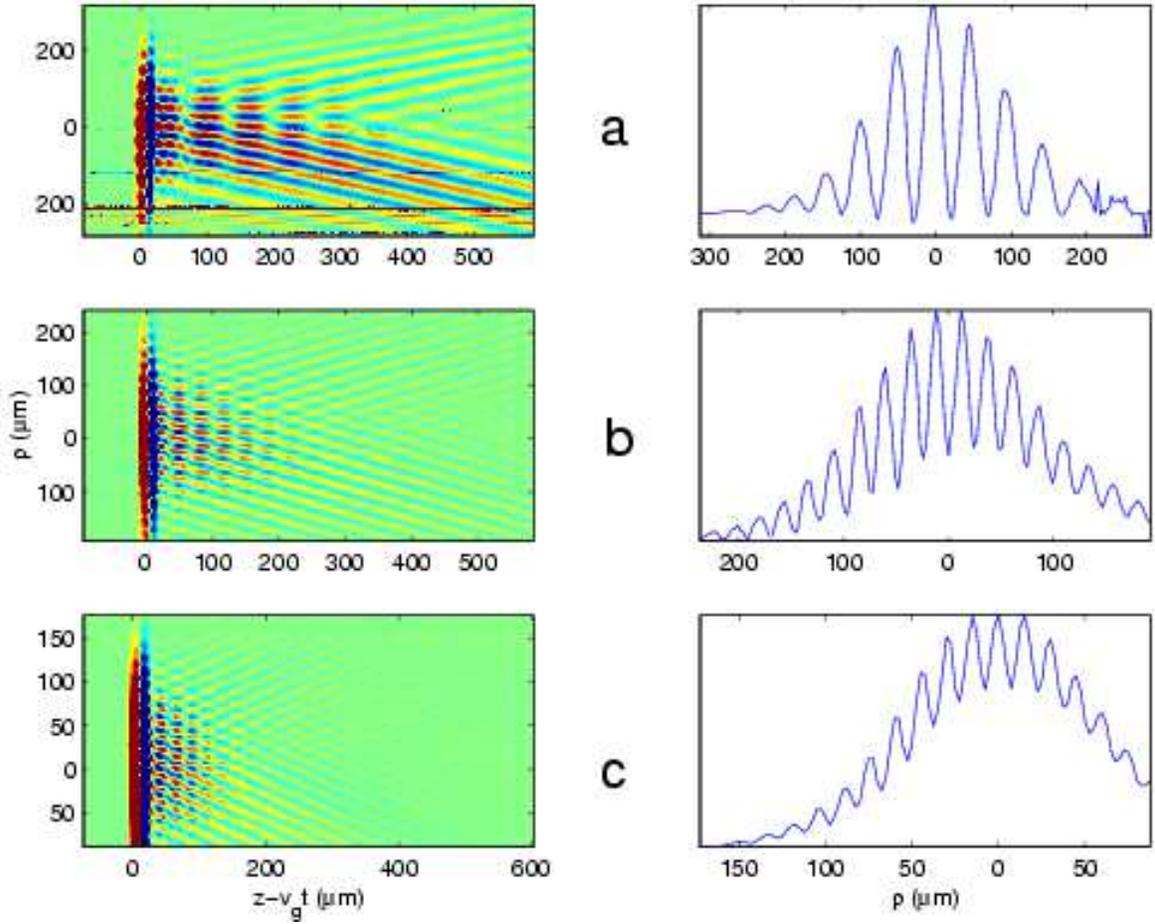}}}
\caption{ Results of transient grating experiments in LiTaO$_3$. Left:
Derivative of the polariton field as a function of time
delay and position of the pump lens. Right: Cross sections at zero
time delay, showing the intensity grating's period. The phase matched
frequency is (a) 1 THz (b) 2 THz (c) 3 THz. Damping is higher at higher
polariton frequencies. }
\label{gratexptfig}
\end{figure*}

We adapted the setup described previously to perform experiments with two
pump pulses, creating a periodic dipolar distribution in LiTaO$_{3}$, and
used frequency shift detection to map the polariton field. The pump pulse
was split into two equally intense pulses which were focused by a single
lens and crossed inside the sample at an angle $\alpha $. Using a beam waist
such that  $w<<v_{g}\tau /\sin \alpha $, we avoided the so-called ``pancake
effect,''\cite{crimmins02} a shrinking of the overlap between weakly-focused
pulses due to their short duration. In our experiments, we did not have to
measure $\alpha $, because the grating period was directly obtained by
measuring the coherent artifact as a function of spatial position. As
for the point geometry, we mapped the polariton field due to this pump
configuration as a function of $z-v_{g}t$ and $\rho $. The results in Fig.~%
\ref{gratexptfig} show excellent agreement with the theoretical predictions.

\subsection{LiTaO$_3$: Phonons and Polaritons}

\begin{figure}[tbp]
\centerline{\includegraphics{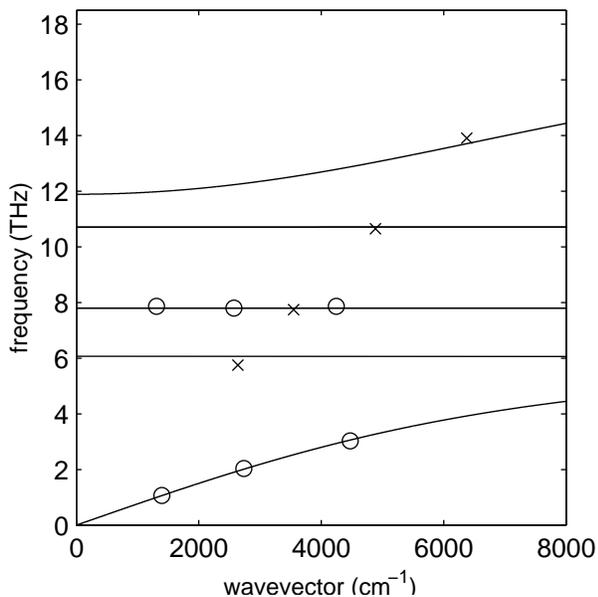}}
\caption{Summary of results in LiTaO$_3$. Circles indicate frequencies
measured using the grating technique, with wavevector given by $%
k_x/\sin\theta$, where $k_x$ is the wavevector of the grating and $\theta$
is the Cherenkov angle. Crosses indicate measurements with a planar
geometry, with wavevector given by $\Omega/v_g$. The solid lines give the
polariton dispersion according to the infrared measurements of Ref.~\onlinecite{barker70}.
}
\label{exptlitafig}
\end{figure}

LiTaO$_{3}$ has a particularly rich history in the ISRS literature.\cite
{bakker92,bakker93,wiederrecht95,brennan97,perry97,bakker98,bakker98b,romero99,koehl99,koehl01a,crimmins02}
In this section we compare our data to those of other reports. There are two
main points of contention in the literature: the origin of
frequency-dependent damping effects and avoided crossings at low frequency.
Our results for the polariton dispersion are summarized in Fig.~\ref{exptlitafig}.

As previously discussed for the transient grating geometry, see below Eq.~(\ref
{pmeq}), the Cherenkov angle appears in the expression for the polariton
wavevector. Using the correct expression, our data fall almost exactly on
the dispersion relation of the lowest branch $A_{1}$ polaritons inferred
from the IR data of Barker et al.\cite{barker70} Recent ISRS papers have
begun to take this into account as well.\cite{hebling02,crimmins02} The
grating technique is also sensitive to non-dispersive phonons, and the
signal at 7.8 THz roughly matches an $A_{1}$ phonon peak seen in Raman
spectra.\cite{Raptis88} Scans in the planar geometry, with wavevector
given by the group velocity of the pulse, are also shown in Fig.~\ref
{exptlitafig}. Here, we are sensitive to even more phonon branches,
including a higher-branch polariton above 12 THz. Note that higher branches
were also observed in a recent heterodyne experiment.\cite{crimmins02}

Our measurements of the polariton damping rate $\gamma $ by the grating
method,\cite{damping} agree very well with those reported by others: $\gamma
=2.4$ cm$^{-1}$ at 1 THz, $\gamma =4.4$ cm$^{-1}$ at 2 THz, and $\gamma =7.5$
cm$^{-1}$ at 3 THz. Large damping, peaked at 2.7 THz, was reported by Auston
and Nuss in their original experiment.\cite{auston88} They suggested that
this was due to multi-phonon absorption. In their homodyne transient grating
experiment, Wiederrecht et al \cite{wiederrecht95} measured increased
damping peaked at 2.76 THz and 4.38 THz, and attributed this to
strain-induced coupling to heavily damped $E$-symmetry modes at the same
frequencies. Very recently, Crimmins, Stoyanov, and Nelson measured the
damping rate using heterodyne detection and suggested that the damping is
due to quadratic coupling between the phonon-polariton mode at 2.1 THz and
the lowest-lying $E$-mode at 4.2 THz.\cite{crimmins02}

The inelastic neutron scattering data of Cheng et al\cite{cheng93} supports
the idea that the damping is due to anharmonicity, more precisely, to a
process involving the simultaneous creation and destruction of two different phonons.
Close to the observed peaks at 2.76 and 4.38 THz, the phonon dispersion
shows TO and LA (TA) branches for which the frequency difference at the zone
edge, where the density of states is large, is 2.6 THz (4 THz). In this context, we note that decay by two-phonon
difference modes has been seen in many materials, including ZnTe.\cite
{schall01} Furthermore, many spontaneous Raman studies\cite{penna76,Raptis88}%
, including one on our particular crystal,\cite{wahlstrandunpub} show a
broad peak at 2.55 THz, consistent with a second-order feature. An
alternative explanation for polariton broadening, supported by the fact that
the TA and LA branches at the zone edge are at 3.2 and 4.6 THz, respectively,%
\cite{cheng93} is coupling to high-wavevector phonons due to
disorder-induced breakdown of momentum conservation. Given that LiTaO$_{3}$
is often non-stoichiometric, defect-related processes may contribute
strongly to the damping.

Another anomaly reported in LiTaO$_{3}$ is an extra avoided crossing in the
polariton dispersion at 0.9 THz.\cite{bakker92,bakker93,bakker98} Similar
avoided crossings were also reported for LiNbO$_{3}$,\cite{bakker94} and
increased absorption was observed at the corresponding frequencies in
stimulated Raman gain measurements.\cite{qiu95} The avoided
crossings were assigned by Bakker and co-workers\cite{bakker98} to quantum
beats between levels of an intrinsic anharmonic potential. However, our
work on LiTaO$_3$, as well as a recent heterodyne ISRS experiment,\cite{crimmins02} show
no sign of an avoided crossing at 0.9 THz. The observation of several
low-lying modes in a CARS study of MgO-doped LiNbO$_{3}$\cite{schwarz98}
suggests that the extra modes are due to defects. If this were the case, the
reported discrepancies for LiTaO$_3$ may simply be due to differences in the concentration
of defects for different samples.

We take issue with the intrinsic anharmonicity proposed by Bakker et al\cite
{bakker98} to explain these extra modes. Unlike in molecules, lattice
anharmonicity in solids leads to decay, because anharmonic coupling involves
a continuum of modes. Hence, unless the anharmonicity is strong enough to
support 2-phonon bound states, it cannot introduce new frequencies. To the
best of our knowledge, no such bound states have been observed in LiTaO$_{3}$
or LiNbO$_3$.

An unexplored area is the role of anisotropy in LiTaO$_3$. As briefly
discussed earlier, for frequencies approaching the lowest $E$-phonon
frequency, the index of refraction becomes extremely anisotropic. Anisotropy
is known to have an important effect on CR\cite{zrelov70} but as far as we
know, it has been completely ignored in the ISRS literature. For a point
charge traveling through a uniaxial material in a direction perpendicular to
the optical axis, both ordinary and extraordinary waves are emitted, with
the extraordinary waves polarized at a finite angle to the optical axis.
Assuming this is also true for a dipole oriented along the optic axis, it
should have a large effect on the CR pattern.

Finally, we note the observation of wavevector overtones in transient
grating experiments using amplified pulses, which Nelson et al\cite
{brennan97,romero99} ascribed to lattice anharmonicity. Since polariton
nonlinearity is primarily due to phonon anharmonicity and the polaritons in
question are mainly light-like, we consider this interpretation to be
questionable. Instead, we propose that the higher-order diffraction peaks
simply reflect a non-sinusoidal grating due to higher-order optical
nonlinearities. This is an area which needs to be explored further.

\section{Conclusions}

In summary, we have used ultrafast optical pulses to generate and image
coherent polariton fields in ZnTe, GaP and LiTaO$_3$. For single and
multiple pulses of arbitrary shape and, in particular, for the
transient-grating geometry, we have shown that the polariton field can be
calculated by convolving the slowly-varying pulse envelope with the
Cherenkov field due to a dipole. The standard phase matching arguments can
also be explained by applying ideas from the theory of Cherenkov radiation.
Results for point-like and planar sources reveal important differences
between the superluminal and subluminal regimes, especially for the
Cherenkov emission by a plane of dipoles which can only occur in the
subluminal case.

\begin{acknowledgments}
We thank G.~Narayanasamy for assistance in the early stages of the experiments and T.~E.~Stevens for useful discussions.
We are grateful to M.~DeCamp, C.~Herne, and J.~Murray for providing the ZnTe samples, and J.~Deibel for the LiTaO$_3$ sample.
This work was supported by the AFOSR under contract F49620-00-1-0328 through the MURI program, and by the NSF FOCUS Physics Frontier Center.
\end{acknowledgments}

\bibliography{pol}

\begin{thebibliography}{60}
\bibitem[{dam()}]{damping}
\bibinfo{note}{Due to the complexity of the Cherenkov field pattern, extreme
  care must be taken in extracting damping from ISRS data. The values quoted
  are from measurements using a large transient grating, in which case spurious
  edge effects are minimized.}
\end{thebibliography}

\end{document}